\documentclass[review]{elsarticle}

\usepackage{lineno,hyperref}
\modulolinenumbers[5]

\journal{Journal of \LaTeX\ Templates}
\usepackage{amsfonts}
\usepackage{amssymb}
\usepackage{amsmath}
\usepackage{mathtools}
\usepackage{verbatim}
\usepackage{soul}
\usepackage{color}
\usepackage{graphicx}
\usepackage{subfig}
\usepackage{float}

\begin{document}

\begin{frontmatter}

\title{Micromechanical model for sintering and damage in viscoelastic porous ice and snow. Part I: Theory.}
\tnotetext[mytitlenote]{Validation and calibrition of the model presented in this paper are described in the second part of this paper.}

\author{B. Wendlassida KABORE }
\ead{brice.wendlassida@gmail.com}
\author{Bernhard PETERS }
\address{University of Luxembourg, 2, avenue de l'Université, 4365 Esch-sur-Alzette}



\cortext[mycorrespondingauthor]{B. Wendlassida Kabore}


\begin{abstract}
Ice and snow have sometime been classified as a viscoelastic or viscoplastic material according to temperature, strain rate, pressure and time scale. Throughout experimental studies presented in the literature, it has been observed that at very low temperatures or high strain rate, porous ice and snow exhibit brittle behavior, but experience high viscous and plastic flow at temperatures closed to the melting point and low rates. 
At the macroscopic level nonlinearity is not necessarily attributed to material level permanent changes or yielding but mainly to micro cracks, porosity collapse and crack propagation. This paper attempts to address this complex behavior with a full microstructure based model.  

\end{abstract}

\begin{keyword}
Microstructure \sep viscoelasticity \sep particle method \sep fracture \sep beams
\MSC[2010] 74M25\sep  99-00
\end{keyword}

\end{frontmatter}

\section{Introduction}


  Several engineering experts attempted to simulate snow behavior for different applications such as movies, avalanche protection and prediction, ski, tires, civil infrastructures. Snow is an heterogeneous media. Simulating and modeling its mechanical response at large scale require a detailed analysis. Such analysis can be done in three steps. The first step concerns the design and validation of mathematical model of small length and time scales describing the grain scale behavior and including most of relevant micro-mechanical processes. Static and dynamic properties including frequency and rate dependency are studied and modeled.
secondly the small length and time scales model is used to link  numerical simulation with meso-scale mechanical behavior of laboratory scales representative volumes. Third a macro-scale model is set with less complexity for simulating the large structure response. For most material, the first step can be skipped as only homogenized version of the true representative volumes is considered. However given the structure dependent mechanical response and the fracture properties of snow it is difficult to reach realistic modeling without this step. 
The conditions of interest in this study are slow, rapid and large deformation for which thermodynamic state of the material is important. The main objective of the presented model is to provide a reliable simulation tool for investigating the non-linearity,  rate, load and temperature dependent mechanical response of snow with complex boundary and loading conditions in engineering. Two phases of snow are considered: the granular phase in which snow is an aggregation of contacting granules and the continuum phase a solid porous ice media. In the granular phase, ice grains are free to move and particles are characterized trough frictional and inelastic contact. In the continuum phase grains are bonded in a melting and re-crystallization cycle and form together a solid structure. The difficulty lay in capturing the geometry with accuracy and its evolution caused by slow or rapid processes at different time scales. The slow processes known as metamorphism include a melting and appreciable flow of a liquid phase followed by crystallization or a diffusion sintering driven by energy minimization. These processes are classified into three categories according to the temperature gradient: the equilibrium metamorphism (Equi-temperature) that turns the initial crystal into rounded, kinetic metamorphism that creates faceted grains and melt-freeze metamorphism for large round grains ~\cite{SnowSnowpack,Cresseri,Adams2008b}. The temperature gradient is behind formation of layers of different strength. A weak layer lead to high risk of avalanche in a snow pack. The rapid processes include mechanical rearrangement of grains, fracture and pressure sintering caused by external forces. This paper presents the short time scale behavior of snow with exception of wet snow. A coupled micro-beam lattice model and discrete particle model is proposed.



\section{Modeling sintering effect of in snow dynamics}
\subsection{Free sintering and pressure sintering}
 Snow behavior is characterized by fracture, creep, and the dimensions of its constrictions (bond between grains) created through sintering. The bonding and adhesion of ice have been largely studied in the past. Some effort were made to characterize the adhesive forces between bonded ice particles at different sintering time and temperature. When two spherical ice grains were brought in contact they became quickly bonded   \cite{Faraday1859NoteRegelation,Nakaya1954,D.Kingery1960,Hobbs1964,Gubler1982,Kuroiwac,Szabo2007}. This phenomenon was first documented by Faraday \cite{Faraday1859NoteRegelation} in 1850 who noticed that two ice blocks became one when brought into contact. 
Most of experiments performed at large time scale (few minutes to days) showed that the neck growth can be attributed to mass transport to the contact area. These experiments showed a temperature dependent growth with a very rapid growth temperatures close to the melting point. However, separate studies show that sintering also happen at shorter time scale when the contact interface is under pressure. The shape of the bond are different, given the fact that the diffusion sintering is mainly caused by mass transport to the neck while the pressure sintering happen by increasing contact area through deformation (figure 1). Processes in both pressure and diffusion sintering are largely attributed to the presence of a quasi-liquid layer on the ice boundary \cite{Weyl1951,Nakaya1954,Jellinek1967} \cite{Fletcher1968} or the melting of the interface \cite{Gubler1982}. The thin quasi-liquid layer is almost always created at the surface of ice or water in an attempt to reduce the free energy, as surface ions change their electron distribution 
\cite{Weyl1951}. The thickness of the liquid layer was estimated to be around 10 nanometers \cite{Weyl1951} ie. Fletcher measured $1$ to $4$ $ nanometers$ \cite{Fletcher1968} between $-8 $ and $-1^o C$  and Jellinek $90$ $nonometers$ at $-1.8^oC$ \cite{Jellinek1967}.
\begin{figure}[H]
\begin{center}
\includegraphics[scale=0.06]{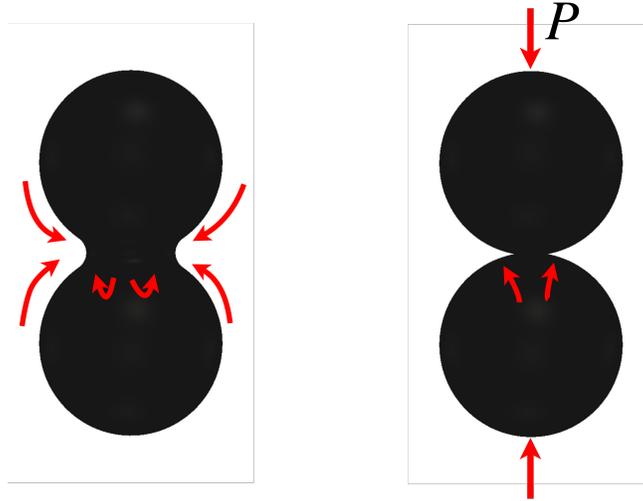}
\caption
{ Geometry and mass transport in free (diffusion) sintering (left) and pressure sintering}
\label{fig:f2}
\end{center}
\end{figure}
Some quantitative studies presenting the effect of pressure in the initial stage of the sintering have emerged in the last decades \cite{Gubler1982,Szabo2007}. The pressure sintering is mainly due to regelation, a melting and freezing process  by variation of the melting point according to the pressure. Ice regelation was first documented in a quantitative study carried by J. Thomson and W. Thomson  \cite{Thomson1850} who called into question Faraday's quasi-liquid layer theory. Regelation is caused by an increase of pressure followed by a decrease attributable to variation of external forces, relaxation, progressive increase of surface area, local melting of asperities .... When passing from compression to tension, the liquid portion previously melted by pressure plus the portion from the quasi-liquid layer accumulated at the contact area freeze back and resist to the tension. Some portion of liquid for instance melted of asperities and liquid layer freeze while still under compression when the contact interface is widened by time dependent deformation. Experiments have shown that after fracture of the bond, the new surfaces of both particles recover.

\subsection{Bond growth and evolution of porous structure}

In the light of the above assumptions, the load carrying capacity $(f ^b_{max})$ of bonds created by pressure sintering can be calculated in two components. A first component $f _0 ^b$ being pressure independent is composed of resistance arising from rapid freezing of quasi-liquid layer at the contact and quasi-instantaneous capillary attraction \cite{Faraday1859NoteRegelation}. This component is present for both pressure and diffusion sintering. The second component $f ^b(p,t)$ which is pressure $(p)$, time $(t)$ and temperature $(T)$ dependent originates from contact interface being welded by melting and freezing. The later dependents on mechanical properties of ice and the loading condition.
\begin{equation}
f ^b_{max}=f _0^b + f^b(p,t,T)
\end{equation}  
Pressure sintering cannot account for total adhesive forces experienced in ice-ice contact \cite{Jellinek1967}. However for short time scale, or appreciable pressure, the pressure dependent adhesive force is dominant. This was proved in the experiments performed by Nakaya \cite{Nakaya1954}, where cohesion increased more than ten time when the contact force was increased by ten. Since our focus is the short time scale only pressure sintering is considered. Diffusion caused neck growth are not considered throughout the simulation though initial micro-structures can be taken from any stage of the metamorphism. In order to apprehend the mechanical behavior of the complex structures resulting from pressure sintering, the structure of porous ice or snow is represented by discrete particles and bonded virtually by massless cylindrical beams. The ice grains are the particles and have frictional contact properties while the constrictions in the ice matrix are the beams connecting particle pairs \cite{Kabore2018Multi-scaleMechanics}. When the structure loses all its bonds it become granular media and can go back to its porous solid structure if given enough time and pressure to sinter again. 
Stress state at the contact area determines which framework is considered for assessing the mechanical response to external loads. For compressive stresses below the compressive strength, a time dependent bond growth or re-bonding mechanism due to visco-plasticity and melt-freeze mechanism occur. The growth of the bond area $(A_b)$ is directly linked to the viscous and plastic deformation of the ice particles and their equivalent radius $(r_{ij})$ as shown in equation 2.
\begin{equation}
A_b= \psi(u, r_{ij})\\
\end{equation}
For torques, tensile and shear stresses, there is viscoelastic deformation accompanied by quasi-brittle fracture of the bonds. The micro-structure evolution is represented by coupling discrete particle model for the first mechanism and a lattice of Euler-Bernoulli micro-beam with fracture properties for the latter. The previous description applies to solid phase. When there is no bond or negligible bond radius i.e granular phase torques and shear resistance are computed according to frictional contact between particles. 
This conceptual model is consistent with the creep, collapse and flow mechanisms discussed in \cite{Bader1962}. 


\subsection{Viscoelasticity of Ice and snow}

Under specific modeling constrains and for the sake of simplicity ice can be considered elastic. However, the conditions under which ice exhibits pure elasticity are so restricted and can hardly be met in real life \cite{Voitkovskii}. Attempts to characterize ice by means of young modulus under static experiments resulted in wide range of values with one order of magnitude difference \cite{HomerT.Mantis}. Measurement using high frequency are considered more reliable.
Several researcher concluded that plasticity can be observed under any stress and the elastic limit is usually assumed to be null, similar to viscoelastic materials. The similarity of ice and viscous fluid has been pointed out by observation of glaciers flow and ice creep in laboratory. Figure 2 shows laboratory measurement of secondary creep rate of ice \cite{Barnes} according to the stress.
\begin{figure}[H]
\includegraphics[scale=0.5]{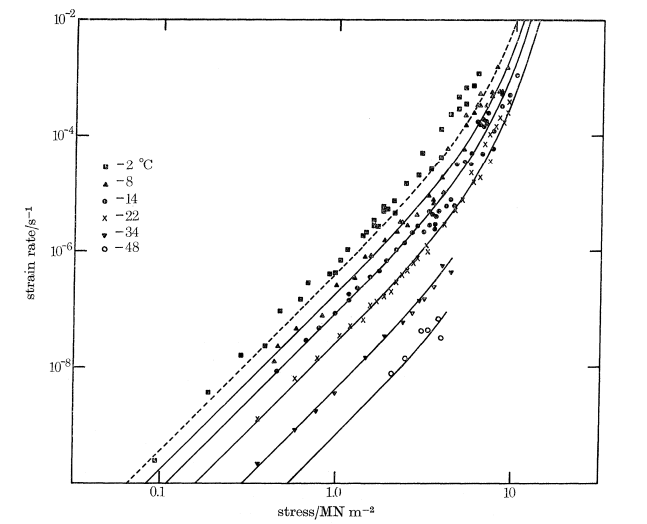}
\caption
{Stress dependence for of the creep rate Barnes 1971 \cite{Barnes}}
\label{fig:f2}
\end{figure}
Pure elastic behavior of ice can be obtained at high strain rate\cite{Mellor} and hydrostatic compression of a single crystal \cite{Voitkovskii}. The main cause of viscous behavior in ice is its crystalline structure. Ice is generally found under polycrystalline form which is composed of single crystalline plates the order of magnitude of $10^2$ micrometer thick \cite{Voitkovskii}.
In deed a single ice crystal can be characterized as elasto-plastic. Since polycrystalline ice is formed by plates the deformation is also dependent to the orientation of the crystals according to the applied stress.
The deformation and creep of polycrystalline ice is characterized by basal dislocations gliding of ice crystals along the basal planes \cite{Barnes}. For randomly oriented crystals, deformation is accompanied by bending and shearing of the crystals. Since ice  is often found at temperature closed to its melting point, its behavior is dominated by creep flow. The creep of ice is divided into tree stages. First stage is the transient creep characterized by a high strain rate that decreases rapidly to a limit. The limit is the secondary or steady state creep that is followed by tertiary creep which leads to failure. The decrease of strain rate during transient creep is about 2 order of magnitude \cite{Castelnau2008}. Most of the strain caused during transient creep is recoverable. The transient creep plays a significant role in ice and snow dynamics.
Therefore for limited time, ice can be considered as a non-Newtonian viscoelastic fluid ~\cite{Bader1962}.


\section{Integrated model for ice and snow dynamics}



The model developed in this paper is a coupling between discrete particles method and damageable Euler-Bernouli beams lattice of grain scale to yield viscoplastic behavior at the macroscopic level. The compressive behavior of ice grains is described by a linear viscoelastic contact model with different creep mechanisms. Despite the fact that nonlinear models have been used to include secondary creep, it has been found that secondary creep is rarely obtained under limited time of observation and that tertiary creep usually happens during the transient creep for high stresses \cite{Duval1974}. Under shear, tension, bending and torsion, we use a viscoelastic quasi-brittle beam model for bonded grains and an elastic perfectly-plastic law for detached grains.

\subsection{Elasticity and creep mechanics for compressive loads}
Using Boltzmann principle and assuming linear viscoelasticity of ice, the displacement can be divided into three independent parts $u_e, u_v, u_{ve}$ (figure 3) in the displacement-time curve for constant loads similar to a Burger's material.
\begin{figure}[H]
\begin{center}
\includegraphics[scale=0.25]{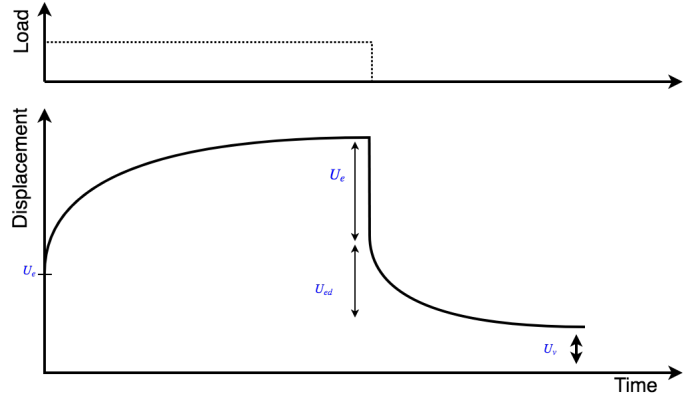}
\caption
{ Displacement-time curve describing vsicoelastic response}
\label{fig:f2}
\end{center}
\end{figure}
The first part of the displacement $u_e$ represents the amount of displacement attributed to instantaneous elastic response. After the load is removed, $u_e$  vanishes almost immediately. Then, a delayed displacement $u_d$ is gradually recovered. Finally, long after the load removal, a permanent displacement $u_v$ representing a Newtonian flow remains. This behavior usually represented by Burger's constitutive model is a series combination of Kelvin and Maxwell models  (figure 4). $u_v$ and $u_e$ are calculated from the Maxwell element and $u_{ve}$ from the Kelvin element. 
\begin{figure}[H]
\begin{center}
\includegraphics[width=8.00cm]{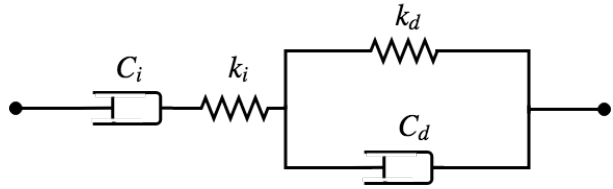} 
\caption
{  Diagram of a four parameter burger material model}
\label{fig:f2}
\end{center}
\end{figure}

For the compressive behavior, we consider rate and loading history dependent equation of Burger's material \cite{Shapiro1997,Yosida,Bader1962} in the following force-displacement relation:
\begin{equation}
f^c + \big [\frac{c_d}{k_d} + c_i (\frac{1}{k_d} + \frac{1}{k_i}) \big ]\dot{f^c} + \frac{c_d c_i}{k_d k_i} \ddot{f^c} = c_i \dot{u} + \frac{c_d c_i}{k_d} \ddot{u}
\end{equation}
Where $c_i$ and $k_i$ are the instantaneous viscosity and stiffness constants, $c_d$ and $k_d$ the delayed viscosity and stiffness constants, $\dot{u}$ and $\ddot{u}$ are the first and second derivatives of the displacement. The instantaneous stiffness can be obtained from the young modulus if measured at high strain rate. \\
The equation $(1)$ is numerically solved for the normal force $f^c_n$ using the central finite difference scheme in the following order \cite{Chen}:

\begin{equation}
\begin{cases}
A=1+\frac{k_d \Delta_t}{2 c_d }; \quad B=1-\frac{k_d \Delta_t}{2 c_d }\\
C=\frac{\Delta_t}{2 c_d A }+\frac{1}{k_i}+\frac{\Delta _t}{2 c_i}; \quad D=\frac{\Delta_t}{2 c_d A }-\frac{1}{k_i}+\frac{\Delta _t}{2 c_i}\\
f^c_n(t+1) = [u+u_d*(1-B/A) - f^c_n(t)*D]\frac{1}{C}\\
u_d(t+1)= \frac{1}{A}\bigg [ B u_d(t) +  \frac{\Delta_t}{2c_d} \big [f^c_n(t+1)+f^c_n(t) \big] \bigg ] \\
\end{cases}
\end{equation}
The same procedure is applied to determine the tangential force $f^c_t$ in the absence of bonds by converting each of the four parameters into transverse their transverse values:
\begin{equation}
P'=\frac{P}{2(1+\nu)}\\
\end{equation}
$\nu$ being the poison ratio.
\subsubsection{Creep and Dynamical response}

The four parameters in (3) can be obtained trough creep test with a compressive impulsion :
\begin{equation}
f^c=f_0 H(t)
\end{equation}
\begin{equation}
H(t)=\begin{cases}
         0  & \quad \text{if } t < 0\\
     1   & \quad \text{if } t\leq 0 \\
  \end{cases}
\end{equation}
The response described in (3) can be reduced to the following using Laplace transformation :
\begin{equation}
u=f[\frac{1}{k_i} + \frac{t}{c_i} + \frac{1}{k_d} \big (1 - e^{-t \cdot t_r} \big)]
\end{equation}
The displacement in the two sections of the Kelvin element are identical and equal to the total delayed displacement. The total delayed displacement is recovered exponentially over time at a rate $t_r$ called relaxation time: $t_r = \frac{k_d}{c_d}$.
The creep rate is the time derivative of the creep:
\begin{equation}
\dot{u}=f[ \frac{1}{c_i} + \frac{\tau}{k_d}  e^{-t\cdot t_r } ]
\end{equation}

Note that the creep rate is the sum of the transient or primary creep rate $f \frac{t_r}{k_d}  e^{-t\cdot t_r}$ and the steady-state or stationary creep rate $ \frac{f}{c_i}$. The transient creep is always present in ice \cite{Duval1974} and is well suited by the Burger's model. However Burger's model is only suitable for short time scale \cite{DiPaola2013} and poorly fit the long term and steady state creep. The longterm behavior may be captured using a nonlinear maxwell dash-pot. The model is also suited for dynamic behavior and response at much smaller time scale than quasi-static conditions. 

Many cases in engineering include dynamic stress or very short contact time (of order of microseconds) between particles. The response under such conditions can well be described in frequency domain. The equation (3) can be written in frequency domain in algebraic form using Lapace transform:
\begin{equation}
(1 +p_1 s + p_2 s^2)f(s)=( q_1 s + q_2 s^2)u(s)
\end{equation}
so that :
\begin{equation}
\frac{f(s)}{u(s)}=sK(s)= \frac{( q_1 s + q_2 s^2)}{(1 +p_1 s + p_2 s^2)}
\end{equation}
A sinusoidal loading of frequency $\omega$ leads to a phase shifted oscillatory displacement of frequency $\omega$ :
\begin{equation}
f=f_0sin(\omega t)
\end{equation}
\begin{equation}
u=u_0 sin(\omega t + \phi)
\end{equation}
Where $\phi$ is the phase angle.
This can be rewritten in the complex domain :
\begin{equation}
f=f _0 e^{i \omega t}= f _0 e^{i \omega t}
\end{equation}
\begin{equation}
u=u _0 e^{-i\phi}e^{i \omega t}= u^* _0 e^{i \omega t}
\end{equation}

The complex compliance is:
\begin{equation}
g(\omega)=\frac{u_0^*}{f_0}=\frac{(1 +p_1  i\omega  + p_2 ( i\omega )^2)}{( q_1 i\omega + q_2 ( i\omega )^2)}
\end{equation}
The phase angle $\phi$, dynamic compliance $g$, storage $g'$  and loss $g''$  compliances representing respectively the angle between force and displacement axis, the modulus, real and imaginary part of $g(\omega)$ can be expressed as:
\begin{equation}
g'=\frac{1}{k_m} + \frac{k_k}{k_k^2 + \omega ^2 C_k ^2}
\end{equation}
\begin{equation}
g''=\frac{1}{\omega c_m} + \frac{\omega c_k}{k_k^2 + \omega ^2 C_k ^2}
\end{equation}
\begin{equation}
g= \sqrt{g'^2 + g''^2}
\end{equation}
\begin{equation}
\tan(\phi)= \frac{g''}{g'}
\end{equation}
The four viscoelastic parameters can also be obtained by fitting equation 16 and 18 to data from dynamic mechanical analysis or oscillatory shear rheometry.


\subsection{Quasi-brittle fracture and collapse mechanics}
\subsubsection{Bond model}
The collapse in ice matrix is described by damageable bond network representing the  constrictions in the ice matrix. Since constrictions are the weakest sections in the snow volume \cite{Ballard}  they are expected to fail before any damageable stress is reached  in the ice grains. Also, because they present small time to failure and designed for ( shear, tension and relative rotations) loading conditions, the maxwell unit in the model for compression is omitted \cite{Shapiro1997} for bonds. 
On the macroscopic level, the bonds between ice particles represent the majority of the constrictions in a snow mass. Therefore, bonds created trough sintering are approximated by cylindrical beams which sections are approximately equal to the area of the constrictions.
It was observed that elastic-brittle bonds were only suitable for fracture in large structures similar to Linear Fracture Mechanics (LEFM) but have limitations in  small size structures. This issue was addressed recently by many investigators ~\cite{Kim2008,Kim2009,Tarokh2014,Galouei2015a,Liu2018} who used bonded particles to analyses fracture of small and medium size structures of quasi-brittle materials. Some softening laws have been introduced to redress the overestimation of liberated kinetic energy after fracture of brittle bonds. These laws include bilinear \cite{Kim2009} exponential \cite{Liu2018} cohesive residual strength beyond the yield point of the material. Furthermore the brittleness in numerical simulation is proportional to the chosen particle size \cite{Tarokh2014}. The exponential softening law is in the form of :
\begin{equation}
f=f_y e^{-\frac{G_f}{\tau} (u-u_l)}; \qquad
G_f=\int_{0}^{\infty} {f d u_\Delta}
\end{equation}
Where $u_\Delta=u-u_l$, $G_f$ is the fracture energy, $f_y = A_b \tau $ is the limit force for a specimen with strength $\tau$.
\begin{figure*}[h]
\begin{center}
\includegraphics[scale=0.32]{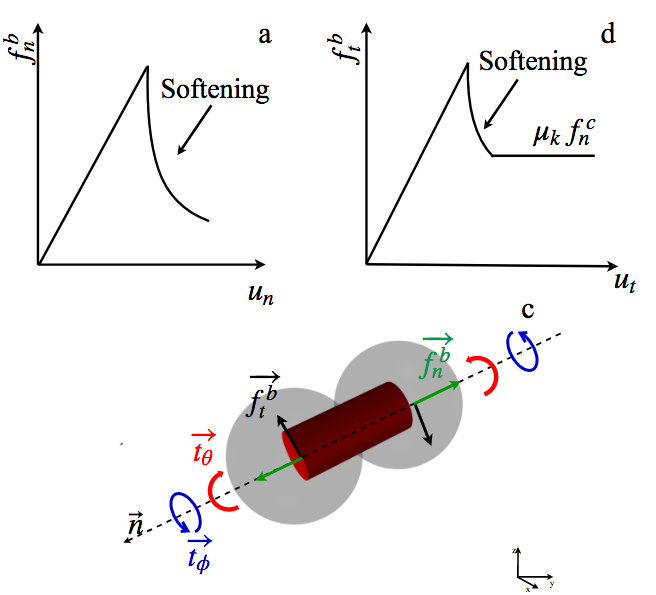}
\caption
{Bonding beam  (a) Force-displacement law in tension, (b) shear with for constant normal compressive force $f_n^c$ (b) Loading conditions of the bond : bending torque $t_\theta$, torsion torque $t_\phi$, tensile displacement $u_n$, shear displacement $u_s$}
\label{fig:f2}
\end{center}
\end{figure*}
The general equation of a homogeneous beam under dynamic load can be formulated as a function strains and distortions using Euler-Lagrange equations :
\begin{equation}
\begin{cases}
EA_b\frac{\partial^2 u_n	}{\partial x^2} + \eta A_b \frac{\partial^2 u_n}{\partial t\partial x} + f_n^b= \rho A_b \frac{\partial^2 u_n}{\partial t^2}\\
GJ\frac{\partial^2 \phi	}{\partial x^2} + \eta J \frac{\partial^3 \phi}{\partial t\partial x^2} + t_\phi= \rho J \frac{\partial^2 \phi}{\partial t^2}\\
EI\frac{\partial^4 u_t	}{\partial x^4} + \eta I \frac{\partial^5 u_t}{\partial t\partial x^4} +\frac{\partial t_\theta}{\partial x}     + f_t^b= \rho A_b \frac{\partial^2 u_t}{\partial t^2}
\end{cases}
\end{equation}
\begin{equation}
\begin{cases}
\theta = \frac{\partial u_t	}{\partial x}= \Delta t (n \times \omega _i - n \times \omega _j)\\
\phi=\Delta t (n \cdot \omega _i - n \cdot\omega _j)n\\
\end{cases}
\end{equation}

Where $f_n$ and $f_t$ are the unbalanced normal and transverse loads, $u_n$ and $u_t$ the normal and transverse displacement, $\phi$ is  the relative spin along the normal direction or twisting angle and $\theta $ the bending angle. $\omega _i$ and $\omega _j$ are the angular velocities. The parameters $E$, $G$, $\eta_n$ and $\eta_t$ are the Kelvin elastic moduli and viscosities of ice in normal and shear direction.
The Saint-Venant assumptions is considered for the twist about neutral axis or shear center (torsion) neglecting the warping torsional moment. Under compression, the resulting bonding beam start thickening and the bond area $A_b$ increases. When loaded, the resistance forces and torques of the bond are calculated as follow :

\begin{equation}
\begin{pmatrix}  f_n^b \\ f_t^b \\ t_\phi \\ t_\theta \end{pmatrix}=\begin{pmatrix}
\eta_n A_b \frac{\dot{u}_n}{ l_b} +EA_b \frac{u_n}{l_b}  \\
\eta_t A_b \frac{\dot{u}_t}{ 2r_b} +GA_b \frac{u_t}{2r_b}  \\
\eta_t \frac{\pi r_b^4\dot{\phi}}{2l_b}+ G\frac{\phi  \pi r_b^4}{2 l_b} \\
\eta_n \frac{\pi r_b^4\dot{\theta} }{4l_b}+ E\frac{\pi r_b^4\theta }{4l_b}  
\end{pmatrix}
\end{equation}

After fracture fracture of a bond, softening functions are used to dissipate energy until the fracture energy is considerably released :
\begin{equation}
\begin{pmatrix}  f_n^b \\ f_t^b \\ t_\phi \\ t_\theta \end{pmatrix}_f=\begin{pmatrix}  f_n^b & f_t^b & t_\phi & t_\theta \end{pmatrix}_{limit}
\begin{pmatrix}  e^{-\frac{G_f}{\tau_n} (u_n-u_{nl})} \\  e^{-\frac{G_f}{\tau_s} (u_t-u_{tl})} \\ e^{-\frac{G_f}{\tau_s} r_b(\phi-\phi_l)} \\  e^{-\frac{G_f}{\tau_n}  r_b(\theta-\theta_l)}  \end{pmatrix}
\end{equation}

 $u_{.l}$, $\phi_l$,$\theta_l$ are respectively the elastic limit of  displacement, twist angle and bending angle. $\tau_s$ and $\tau_n$ are the shear and normal strength.

The length of the beam $l_b$ is the distance between the center of mass of the particles and the bond area is equal to the contact area  :
\begin{align*}
A_b \approx \pi r_{ij} u_n \\
\end{align*}
A local coordinate is used for the pair particles, while the shear force and bending moment are stored in the global coordinate.
The rotation of the local coordinate is taken into account for the shear, and torques. The angle of rotation $\varphi$ between the current normal  $\vec{n_{t+1}}$ and previous $\vec{n_t}$ about an axis $\vec{n'}$:
\begin{align}
\vec{n}'=\vec{n}_t \times \vec{n}_{t+1} \\
\varphi = \arcsin{|n'|}
\end{align}
A rotation matrix in global coordinate is calculated from the rotation in local coordinate using quaternion transformation. The torques and shear forces in previous time step are rotated to the new local coordinate before being updated.

\subsubsection{Quasi-brittle fracture}{
The fracture behavior of ice have been thoroughly studied in the past for different loading rates\cite{Schulson2006,Schulson2001,Schulson2009,Schulson1987}. It was found that ice grain of 1 to 2 mm size showed brittle behavior for strain rates above $10^{-7} s^{-1}$ in tension and above $10^{-3}s^{-1}$ in compression at $-10^o C$. These rates represents the transition zone  between ductile and brittle behavior \cite{Schulson2001}. The transition rates are lower for larger sizes and for compressive loads, it decreases with temperature. The fracture behavior depends on the creep rate of the ice. The ductility at low rates is due to the fact that stress relaxation rate is high enough to inhibit stress concentration and crack growth. 
The failure criterion combining tensile and bending stress of the bond is expressed as follow \cite{Potyondy2004}:
\begin{equation}
\frac{f_n^b}{A_b}+\frac{4 t_t}{r_b A_b} < \tau_n
\end{equation}
In case of shear damage the Mohr-Coulomb criteria is used. The total shear resistance obey the Mohr-Coulomb shear strength. In the Mohr-Coulomb failure criteria, the shear strength is expressed as:
\begin{align}{
\tau_s =\frac{f_n^c}{A_b} \mu_s + C
}
\end{align}
The shear failure criterion combining shear and torsion stresses of the bond is expressed as follow :
\begin{equation}
\frac{f_t^b}{A_b}+\frac{4 t_n}{r_b A_b} < \tau_s
\end{equation}
$\mu_s$ is the coefficient of static friction and $\frac{f_n^c}{A_b} \mu_s$ is the residual strength after bond fracture and is always present in granular snow~\cite{Mellor}. 
This criterion fits well with our modeling, in such a way that the load carrying capacity of bonds are increasing with pressure and time. Since the bonds size are time and temperature-dependent the value of C is: 
\begin{align}{
C =A_b \tau _{ice} = \Psi(f^c_n, t^b, T )  \tau _{ice}
}
\end{align}

Another important parameter in discrete particle model is the ratio between the size of the real material grains and the size of particles used. In addition to fracture poor resolution, large ratio may lead to some innacuracies due to the Hall$ - $Petch effect. The Hall$ - $Petch effect is the phenomenon in which materials are strengthening or weakening when their average grain size are changed. In fact it is observed that as the grain size get smaller  the material exhibit higher strength. The tensile strength and grain size are therefore related through the following expression:
\begin{align*}
\tau _{ice}=\tau _0 + kd^x \\
\end{align*}
For ice at $-10^oC$, $x=0.5$, $\tau _0 =0.6 MPa$, $k=0.002MPa \sqrt[]{m}$ \cite{Schulson2006}.
$k$ is the strengthening coefficient. 


}

\subsection{Friction and flow mechanics}{
After fracture, relative shearing and rotation still lead to viscoelastic and plastic displacements. The plastic displacement can be modeled as a Coulomb friction. Friction forces between particles play a significant role in macroscopic behavior of ice and snow and are major source of energy dissipation.
\subsubsection{Relative sliding}
For shear forces, the resistance is characterized by a static and kinetic friction coefficient $\mu_s$ and $\mu_k$. The static friction is the maximum limit of the residual shear resistance $f_t^c$ described above. 
Once the relative motion between particles become appreciable, the static friction resistance vanishes and is replaced by kinetic friction force. The frictional resistance force $f_t^c$ can be expressed for a normal contact force $f_n^c$ and a tractive force $ f_t$ as:
\begin{align}{
f_t^c =
  \begin{cases}
    f_t ^c     & \quad \text{if }  f_t^c < \mu_s f_n^c\\
    \mu_k f_n^c  & \quad \text{if }  f_t^c\geq {\mu_s f_n^c} \\
  \end{cases}
} 
\end{align}

The kinetic friction is high for dry and rough surface, and low for smooth and wet surfaces. 
Moreover, It was observed that at very low sliding speed(a few centimeters a second) the kinetic and static friction coefficient for snow or ice are close ($10 \%$ difference) ~\cite{Bowden1953}.
Unlike the static friction, the kinetic friction is higher at lower temperature ~\cite{Sukhorukov2013}.  
\subsubsection{Relative rolling and twisting}
Some resistances arise when two ice grains roll against each other. The origin of this resistance may be attributed to the instantaneous cohesion and the elastic hysteresis at the contact area. In this study we consider the following viscoelastic constitutive relation \cite{Iwashita1998} for contact rolling: 
\begin{equation}
\begin{cases}
M^r= -k^r   \Gamma _1 \theta_r  -C_r \dot{\theta}\\
k^r = k^t r_{ij}^2
\end{cases}
\end{equation}

\begin{equation}
\Gamma _1=\begin{cases}
1  & \quad \text{if } |k^r   \theta_r| \leq \mu _r r_{ij} f^n\\
\frac{\mu _r R_r f^n}{|k^r    \theta_r|}  & \quad \text{if } |k^r   \theta_r| > \mu _r R_r f^n\\
\end{cases}
\end{equation}
Where $ \mu _r$ is the coefficient rolling plastic moment. The reduced radius is $r_{ij}=\frac{r_ir_j}{r_i+r_j}$ for two grains and $r_{ij}=r_i$ for a grain $i$ against a wall $j$.
The interparticle torsion for granular phase (not bonded) are neglected.
}
\subsection{Motion integration}

The motion of each particle in the system follows the Newton's second law of motion. For a particle $i$ the translational and rotational motion are updated according to the following equation:
\begin{equation}
\begin{cases}
m_i\dot{v}_i =  \sum_j f_{ij} + m_i g \\
I_i\dot{\omega}_i = \sum_j t_{ij}
\end{cases}
\end{equation}
Where $\sum_j f_{ij}$ and $\sum_j t_{ij}$ are the sum of forces and torques of all interactions between a particle and its neighbors. $m_i$ is the mass of the particle.
The motion of the particle is driven by the unbalanced forces resulting from contact forces in multiple contact environment. The interaction force and torque between two particles $i$ and $j$ are :
\begin{equation}
\begin{cases}
f_{n,ij} =\begin{cases} f^c_{n,ij} \quad \textit{if $\dot{u}_{ij}^n>0$}
\\  f^c_{n,ij} + \zeta_{ij} f_{n,ij}^b \quad \textit{if $\dot{u}_{ij}^n<0$}
\end{cases}\\
f_{t,ij} =\zeta_{ij} f_{t,ij}^b + (1 - \zeta_{ij} )f_{t,ij}^c\\
t_{ij} =  \zeta_{ij} t_{ij}^b + (1 - \zeta_{ij} )t_{ij}^c\\
t_{ij}^c =  f_{t, ij}^c r_{ij}
\end{cases}
\end{equation}
$f_{n,ij}^c$ and $f_{t,ij}^c$ are contact forces in normal and tangential direction, $t_{ij}^c$ contact torques, $f_{n,ij}^b$ and $f_{t,ij}^b$ bond forces in normal and tangential direction, $t_{ij}^b$ is the bond torque. $\zeta_{ij}$ is equal to one if the particles are bonded, and zero if not. 
Interaction forces are calculated according to constitutive model of ice presented above. In discrete element formulation, force-displacement relation is used to describe mechanical behavior in  contrast to continuum mechanics where constitutive laws are often a stress-strain equations.
Displacement of a particle is computed from the overlap (indentation) for each interaction. Hence for two particles the translational and rotational displacement is \cite{Luding2008}
\begin{equation}
\begin{cases}
u_{ij}^n = r_i + r_j -|x_i -x_j| \\
u_{ij}^t =  \Delta t (v_{ij} - n(n \cdot v_{ij}) )\\
v_r ^t=- r_{ij}' \Delta t (n \times \omega _i - n \times \omega _j)\\
v_r^n = r_{ij} \Delta t (n \cdot \omega _i - n \cdot\omega _j)n\\
\end{cases}
\end{equation}
The reduced radius and corrected reduced radius are respectively $r_{ij}=\frac{r_ir_j}{r_i+r_j}$; 
$r_{ij}'=\frac{(r_i-u_{ij}^n)(r_j-u_{ij}^n)}{r_i+r_j- u_{ij}^n}$.

The total displacement is the sum of local displacements on the particle.
We assume that all particles are at rest at the beginning of the simulation, thus no interaction force or residual stress are considered at time prior to the simulation start. Initial displacements usually used to best represent the geometry of sintered parts are removed from all displacement calculation throughout the simulation.


In order to avoid the singularity problem while representing a particle's orientation with three Euler angles ($\alpha$, $\beta$, $\gamma$), a quaternion approach is used.
\begin{equation}
q = \begin{bmatrix}
       \ \cos(\frac{\beta}{2}) \cos(\frac{\alpha+\gamma}{2}) \\
       \sin(\frac{\beta}{2}) \cos(\frac{\alpha-\gamma}{2}) \\
       \sin(\frac{\beta}{2}) \sin(\frac{\alpha-\gamma}{2})\\
       \cos(\frac{\beta}{2}) \sin(\frac{\alpha+\gamma}{2})
     \end{bmatrix}
\end{equation}
the relative angular displacement between two particles is
\begin{equation}
q_{ij}=q_i-q_j
\end{equation}
For numerical stability, we chose the forth order Gear predictor-corrector algorithm \cite{Gear1967,SAMIEI2012} where the motion of each particle is predicted and corrected in the same time step. The prediction is based on Taylor expansion :
\begin{equation}
\begin{cases}
\dot{v}_p(t+1)= \dot{v}(t)+ \frac{\partial ^3 x(t)}{\partial t^3}\Delta t\\
v_p(t+1)= v(t)+  \dot{v}(t) \Delta t + \frac{1}{2}  \frac{\partial ^3 x(t)}{\partial t^3}\Delta t^2  \\
x_p(t+1) = x(t) + v(t) \Delta t + \frac{1}{2} \dot{v}(t) \Delta t^2 + \frac{1}{6} \frac{\partial ^3 x(t)}{\partial t^3}\Delta t^3\\
\end{cases}
\end{equation}
The motion is then corrected at next step:
\begin{equation}
\begin{cases}
\delta=\frac{ f^I(t+1) + mg}{m}-\dot{v}_p(t+1)\\
\dot{v}(t+1)= \dot{v}_p(t+1)+ \delta \\
v(t+1)= v_p(t+1)+ \frac{5}{12}\delta  \Delta t   \\
x(t+1) = x_p(t+1) + \frac{1}{12}\delta  \Delta t^2  \\
\end{cases}
\end{equation}


\section{Verification at grain scale}

The fast sintering experiment consist of putting two spherical ice particles into contact with a given load for a short time and then applying separation force until fracture of the created bond. The applied load, fracture force and sintering time are then recorded. Szabo and Schneebeli performed such experiment by putting two cones of 3mm radius at the tip into contact for different times \cite{Szabo2007}. 
The fracture force $(f_{frac})$ also called sintering force vs time can be translated into strain vs time curve or indentation $d$ vs time using spherical contact mechanics. If the equivalent radius of the two spheres is $R_{eq}$ and the indentation $d$ is known, the bond area can be calculated as follow: 
\begin{equation}
A_b=\pi R_{eq} d = \frac{f_{frac}}{\tau _n}
\end{equation}
Assuming that the tensile strength is constant under the experimental conditions, the indentation can be related to the fracture force as follow :
\begin{equation}
d = \frac{f_{frac}}{\pi \tau _n {eq}}
\end{equation}
\begin{table}[h!]
  \begin{center}
    \caption{Optimal Burgers parameters for short time creep }
    \label{tab:table1}
    \begin{tabular}{l|c|c|c|c|r} 
      \textbf{Temperature} & \textbf{$c_i$} & \textbf{$c _ d$}& \textbf{$k _ i$} & \textbf{$k _ d$} & \textbf{$f_0^b$}\\
      $^o C$ & $MNm. s $ & $MNm. s$ & $MNm$& $MNm$  & $N$\\
      \hline
      -1 & 0.15385e+03 &15.698 & $9.10^3$ &0.30783  &0.08411\\
      -5 & 0.39047e+03 &43.230 &$9.10^3$& 0.53908  &0.06535\\
      -12 & 0.70373e+03 & 81.653 &$9.10^3$ &0.60423  &0.05\\ 
       -23& 1.0444e+03 & 82.50 &$9.10^3$ &1.1561  &0.0298\\
    \end{tabular}
  \end{center}
\end{table}
Under these assumptions, the fast sintering data can be exploited as creep data. Linear extrapolation of measured load dependent sintering force agrees with the existence of non null temperature dependent sintering force when no load has been applied. The load independent portion of the fracture force can be included into equation 42 :
\begin{equation}
d = \frac{f_{frac}-f_0^b}{\pi \tau _n {eq}}
\end{equation}
In this paper, the four parameters of the Burger's model and $f_0^b$ have been obtained by this approach using the damped least-squares (DLS) method. The results of this calibration process are presented in Figure 6 and the parameters are listed in Table 1. The bonds viscoelastic properties are calulated as following : 
\begin{align}
E=(k_i+k_d)d \\
G=\frac{E}{2(1+0.3)}\\
\eta_t=\eta_n =c_i d
\end{align}
\begin{figure}[H]
\includegraphics[scale=0.3]{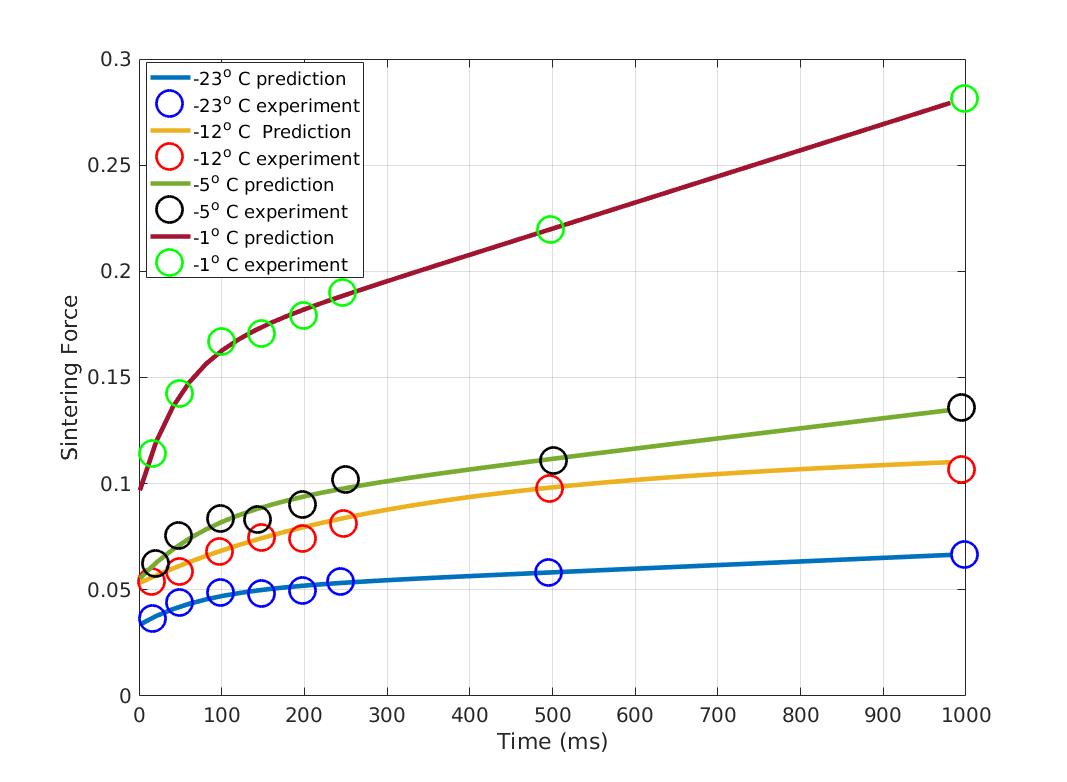}
\caption
{Model prediction vs fast sintering experiments (Szabo and Schneebeli 2007 \cite{Szabo2007})}
\label{fig:f2}
\end{figure} 
In the present model the growth rate of the bond between the particles is linearly dependent on the pressure at the interface. The pressure dependency is in agreement with the experimental data presented by  Szabo and Schneebeli \cite{Szabo2007}. Figure 7 show results of fracture force vs applied pressure of two particles of $3mm$ radius for a sintering time of 250ms. It is worth mentioning that the rate at which the tensile load have been applied for the bond fracture are supposed to allow no appreciable viscous flow.
\begin{figure}[H]
\includegraphics[scale=0.35]{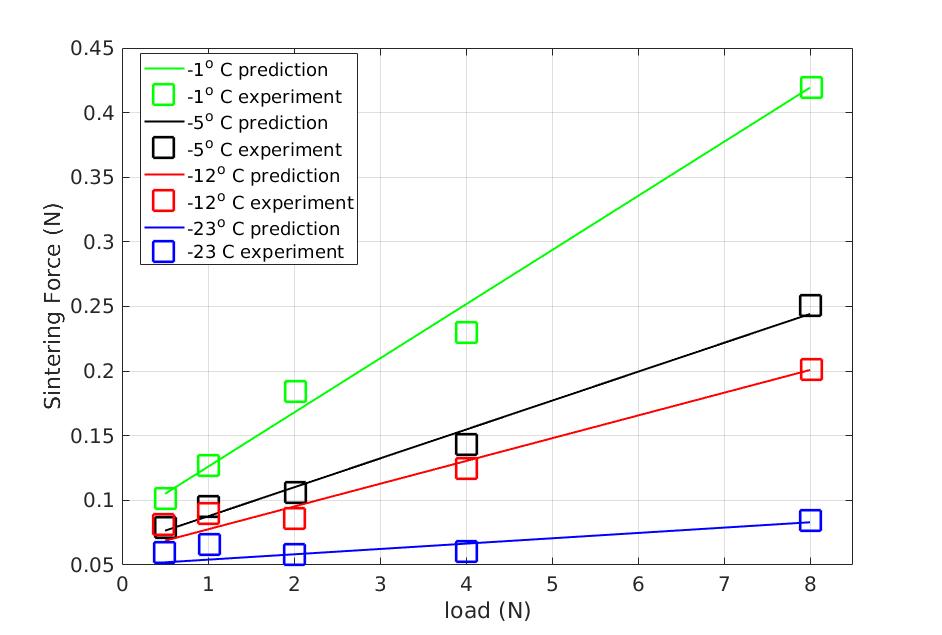}
\caption
{Evolution of sintering force according to applied compressive force compared to experiment Szabo and Schneebeli \cite{Szabo2007}}
\label{fig:f2}
\end{figure}



Time-temperature superposition can be used to establish the relation between viscosity and temperature. For ice, the Arrhenius relation have been used \cite{Barnes,Rist1994,GAGNON1995}.  The Arrhenius type creep rate relation :
\begin{equation}
\dot{\epsilon}=c e^{-\frac{Q}{RT}}
\end{equation}
Barnes proposed a model that describes the secondary creep of ice in a temperature range of $0$ to $-48^o C$ and strain rate ranging between $10^{-9} s^{-1}$ and $10^{-2} s^{-1}$. The activation energy was sugested to be $120 J/mol$ for temperatures above $-8$ and $78 J/mol$ for temperatures below. Other researchers found much lower values for the activation energy $120 J/mol$ for $-40$ to $-20^o C$ \cite{Rist1994} and $101 J/mol$  for temperature range of $-16$ to $-1^o C$ by \cite{GAGNON1995}. The difference in activation energy is believed to come from some liquid at grain boundaries.
Barnes also suggested that by this process the creep rate is supposed to be higher for decreasing grain size. This implies that the creep rate found in individual snow grains are higher than that macroscopic ice leading to even higher macroscopic creep rate of snow. Since densification and grain growth are linked, the creep rate is expected to decrease with increasing density \cite{Barnes,Bader1962}.
In this study, the Williams-Landel-Ferry (WLF) model have been used to establish temperature dependency of viscosity constants:
\begin{align}
c  (T)= a_i(T) c (T_0)\\
a_T=exp\big [\frac{-C1(T-T_0)}{C2+(T-T_0)}\big]
\end{align}
\begin{figure}[H]
\includegraphics[scale=0.5]{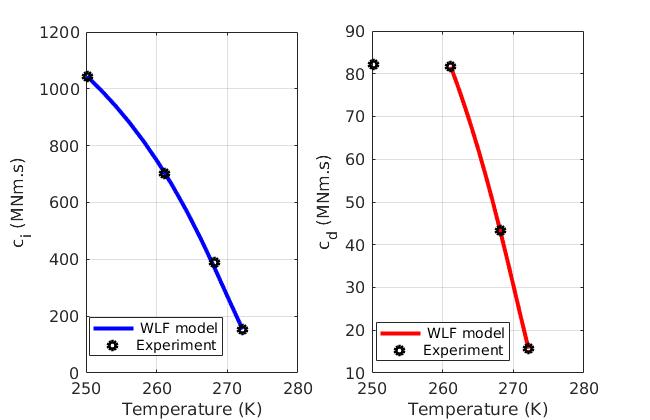}
\caption
{Variation of viscosity constants with temperature}
\label{fig:f2}
\end{figure}
As one can see in figure. 8 the variation of the delayed viscosity is pronounced between $-1$ and $-12^o C$. Below$-12^o C$ no significant increase is found where as the instantaneous viscosity linked to secondary creep still increase.
From the experimental values the following empirical relation have been found:
\begin{align}
c _d (T)= e^ {2.571(T -T_0)/(-6.154+T -T_0)} \cdot c _i (T_0)\\
c _i (T)= e^ {2.586(T -T_0)/(-7.706+T -T_0)} \cdot c _i (T_0)\\
t _r (T)=\frac{c _d (T)}{k_d (T)} = e^ {-1.472\cdot 10^4(T -T_0)/(2.431\cdot 10^5+T -T_0)} \cdot t _r (T_0)
\end{align}
The reference temperature $T_0$ was taken to be $272.15K (-1^o C)$.
\begin{figure}[H]
\includegraphics[scale=0.5]{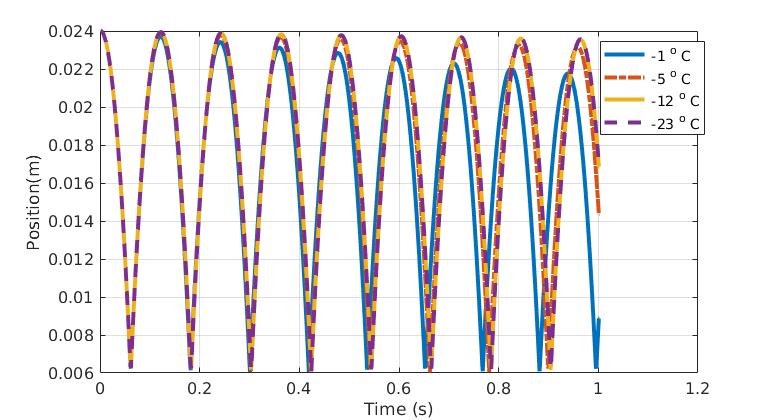}
\caption
{Bouncing particle at different temperature  }
\label{fig:f2}
\end{figure}
Both viscosity constants $c _i$ and $c _i$ are function of temperature and increase with decreasing temperature. The elasticity constant are assumed to be constant with varying temperature and the values are chosen close to the Young modulus. The figure 9 illustrates the influence of temperature on the apparent restitution coefficient. An ice particle falling on ice surface of same temperature show lower restitution coefficient at higher temperature. Although no significant change is found above $-12^o C$.
The fast sintering experiment was also performed by Gubler at different temperature \cite{Gubler1982}. The data extracted from experiments by Szabo and Schneebeli are also in agreement with Gubler's experiments.
\begin{figure}[H]
\includegraphics[scale=0.5]{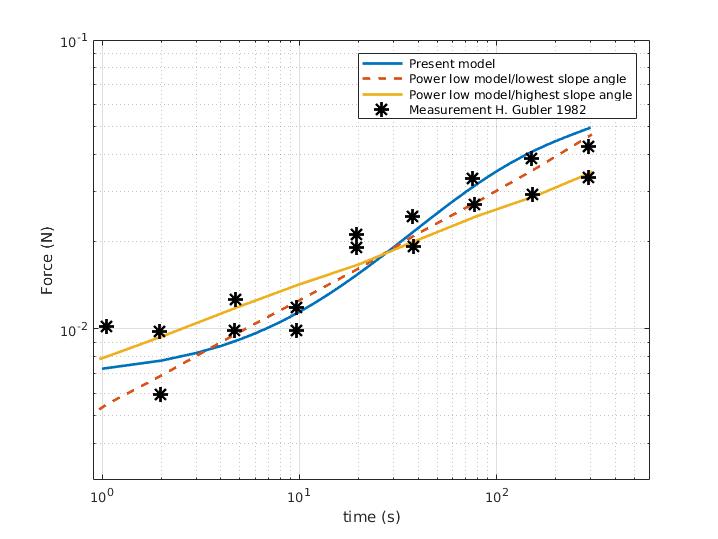}
\caption
{Time evolution of fracture force according to time  at $-10^oC$ }
\label{fig:f2}
\end{figure}

\section{Conclusion}
A conceptual model has been developed to mimic micromechanisms that take place in snow. The macroscopic behavior is governed by grain bonding, de-bonding leading to crack initiation and propagation and all processes that take place in the crack region.
Interactions between ice grains are described using rheological models. Intergranular fracture mechanisms were introduced by the means of quasi-brittle bonds. The model also includes thermo-mechanical description of bond growth. An exponential softening law was used for post-peak behavior of the bonds. 
 
The main features can be summarized as follow:
\begin{itemize}
\item full microstructure is taken in to account for mechanical response;
\item temperature dependent evolution of the microstructure by creep and sintering;
\item size effect in fracture mechanism and rate dependent behavior.

\end{itemize}


\section*{References}
\bibliography{Mendeley}
\end{document}